\documentclass{article}
\usepackage{spconfa4,graphicx}
\usepackage{amsmath,amssymb}
\usepackage{amsfonts}
\usepackage{hyperref,url}
\usepackage{cite}
\usepackage{multicol,multirow,caption}
\usepackage{array}
\usepackage{booktabs,color,comment,flushend}

\DeclareMathOperator*{\argmax}{argmax}
\DeclareMathOperator*{\argmin}{argmin}

\title{Interaural time difference loss for binaural target sound extraction}
\name{
\begin{tabular}{c}
Carlos Hernandez-Olivan, Marc Delcroix, Tsubasa Ochiai,  \\ \textit{Naohiro Tawara, Tomohiro Nakatani, Shoko Araki}
\end{tabular}
\thanks{This work was completed during Carlos Hernandez-Olivan's internship at NTT Corporation. 
Email: carloshero@unizar.es, marc.delcroix@ntt.com. \\
This work was supported by JST Strategic International Collaborative Research Program (SICORP), Grant Number JPMJSC2306, Japan.
}}
\address{NTT Corporation, Japan}
\begin{document}
\ninept
\setlength{\abovedisplayskip}{4pt}
\setlength{\belowdisplayskip}{4pt}

\maketitle
\begin{abstract}
Binaural target sound extraction (TSE) aims to extract a desired sound from a binaural mixture of arbitrary sounds while preserving the spatial cues of the desired sound. Indeed, for many applications, the target sound signal and its spatial cues carry important information about the sound source. Binaural TSE can be realized with a neural network trained to output only the desired sound given a binaural mixture and an embedding characterizing the desired sound class as inputs. Conventional TSE systems are trained using signal-level losses, which measure the difference between the extracted and reference signals for the left and right channels. In this paper, we propose adding explicit spatial losses to better preserve the spatial cues of the target sound. In particular, we explore losses aiming at preserving the interaural level (ILD), phase (IPD), and time differences (ITD). We show experimentally that adding such spatial losses, particularly our newly proposed ITD loss, helps preserve better spatial cues while maintaining the signal-level metrics.

\end{abstract}

\section{Introduction}
\label{sec:intro}

Target sound extraction (TSE) aims at isolating a sound signal of interest belonging to a desired (or specified) sound class from a mixture of arbitrary sounds \cite{ochiai20_interspeech}. For example, TSE could extract the sound of a siren in a recording including also dog barking and car passing. The TSE problem can thus be seen as a generalization of the target \emph{speech} extraction problem \cite{ZmolikovaDOKCY23} to arbitrary \emph{sounds}. Realizing TSE could have many practical applications, such as hearables or hearing aids that can focus on sounds of interest in everyday environments or smart audio post-production systems. 

A typical TSE system consists of a neural network that accepts a sound mixture and outputs the desired sound signal free from other sounds and noise. The network is conditioned on a clue that is used to identify the target sound in the mixture. Several types of clues can be used, such as a one-hot vector identifying the target sound class \cite{ochiai20_interspeech, DelcroixVOKOA23}, audio recording similar to the desired sound \cite{DelcroixVOKOA23, vzmolikova2019speakerbeam, gfeller2021one}, video \cite{ZhaoGRVMT18}, onomatopoeia \cite{OkamotoHYIK22}, the region where to focus \cite{10508743} or multiple clues  \cite{LiQCWYLQZ23}. Here, we focus on TSE conditioned on the target sound class.

Most TSE frameworks focus on single-channel processing \cite{DelcroixVOKOA23, veluri2023real}. However, humans rely on binaural hearing to capture spatial information about the sounds. For example, the interaural level difference (ILD) and interaural time difference (ITD) serve as essential cues to localize sound sources \cite{blauert1997psychophysics,moore1991anatomy}. It is thus essential to develop TSE systems that can extract a target sound while preserving its binaural cues. Binaural processing can be achieved by extending a monaural TSE system to accept binaural input and output binaural signals. This idea was first introduced for speech enhancement \cite{han2020real} and recently applied to TSE  \cite{VeluriICYG23}. We can train such systems using a signal-level loss on each output channel, which guides the model to output left and right signals as close as possible to the references. Such a training loss implicitly pushes the system to preserve spatial cues. However, prior works \cite{han2020real,VeluriICYG23} did not use any explicit spatial loss, which may limit the ability of TSE to recover spatial cues.

This paper investigates whether adding an explicit spatial loss could further improve spatial cue preservation of binaural TSE. 
It is straightforward to define a loss measuring the error of ILD between the estimated and reference signals \cite{tokala2024binaural}, as it only involves computing the difference of the ratio of the norm of the left and right signals, which is differentiable. In contrast, computing the ITD involves finding the position of the maximum in the cross-correlation between the left and right signals. This computation requires the ``argmax'' operation, which is not differentiable, making it challenging to use as a loss. A recent work \cite{tokala2024binaural} has proposed instead a loss measuring the error of the interaural phase difference (IPD), which is related to the time difference. However, optimizing IPD errors may be challenging because of the phase wrapping problem. 

We propose instead an alternative loss, which measures the errors between the cross-correlation coefficients of the estimated and the reference signals.  By abuse of language, we refer to it as ITD loss. The proposed ITD loss is more directly related to the ITD computation than the IPD loss. We thus hope it will lead to better ITD cue preservation. 

Note that a prior works \cite{1661233,tokala2024binaural} have proposed using ILD and IPD losses for binaural \emph{speech} enhancement. However, it is important to explore spatial losses for the TSE problem since it deals with a much larger variety of sounds than speech signals \cite{ochiai20_interspeech,DelcroixVOKOA23,gfeller2021one,veluri2023real}.
To summarize, the contribution of this paper is twofold. First, we propose a new spatial loss that is directly related to ITD cues. Second, we compare three types of spatial losses (ITD, ILD, and the newly proposed ITD losses) for binaural TSE. 
We perform experiments on TSE of sound mixtures containing three to four sounds from 20 sound classes. We use a binaural TSE system as a baseline \cite{VeluriICYG23} and confirm experimentally that adding spatial losses can reduce ITD, ILD, and IPD errors while preserving signal-level metrics. Moreover, the newly proposed ITD loss achieves overall superior binaural TSE performance.

\begin{figure*}[tb]
    \centering
    \includegraphics[width=2.0\columnwidth]{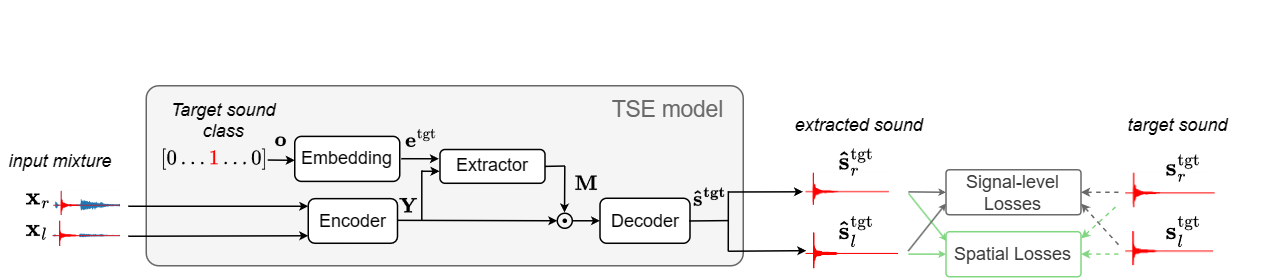}
    \vspace{-2mm}
    \caption{Our proposed binaural TSE system with signal-level and spatial losses.}
    \label{fig:tse}
    \vspace{-5mm}
\end{figure*}
\section{Binaural target SOUND EXTRACTION}
\label{sec:binaural TSE}
In this section, we first introduce the problem statement and describe a baseline binaural TSE model. We then discuss the signal-level and spatial losses we use to train the TSE system. In particular, we introduce our proposed ITD loss in Sec.~\ref{subsec:prop}.

\subsection{Problem Statement}
The objective of TSE is to isolate a sound signal of interest belonging to a desired sound class from an observed signal $\mathbf{x}_m$ consisting of a mixture of sounds defined as:
\begin{equation}
    \mathbf{x}_m = \sum_{i=1}^{I} \mathbf{s}_{i,m},
\end{equation}
where $\mathbf{s}_{i,m} \in \mathbb{R}^{ T}$ is the sound signal of the $i$-th source at the $m$-th microphone, $T$ is the number of samples, and $I$ is the total number of sound sources. 
In the following, we consider binaural recordings and define binaural mixture as $\mathbf{x} = \left[ \mathbf{x}_{l}, \mathbf{x}_{r} \right] \in \mathbb{R}^{T \times 2}$ where  $l$ and $r$ represent the index of the left and right microphones, respectively.
Similarly, we denote by $\mathbf{s}^{\text{tgt}}= \left[ \mathbf{s}_{l}^{\text{tgt}}, \mathbf{s}_{r}^{\text{tgt}} \right] \in \mathbb{R}^{T\times 2}$ the binaural target signal, which corresponds to the desired sound class. 

Binaural TSE aims to recover the binaural signal of the target by preserving the spatial information of the sound.
Therefore, a TSE model follows the expression:
\begin{equation}
    \hat{\mathbf{s}}^{\text{tgt}} = \mathrm{TSE}(\mathbf{x}, \mathbf{o};\theta),
\end{equation}
where $\hat{\mathbf{s}}^{\text{tgt}}$ represents the extracted target sound signal, $\mathrm{TSE}(\cdot)$ is a neural network with parameters $\theta$, and $\mathbf{o} = [0, \ldots, 1, \ldots 0]^{\top}$ is a 1-hot vector representing the target sound class, i.e., it has a value of 1 for the index of the target class and 0 otherwise.  The 1-hot vector has a dimension $O$ corresponding to the number of classes the models can handle.

\subsection{TSE Model}
Our TSE model consists of an encoder-decoder model with a mask estimation network, or extractor, as shown in Fig.\ref{fig:tse}. Unlike conventional single-channel TSE models \cite{DelcroixVOKOA23,veluri2023real}, we use Semantic Hearing \cite{VeluriICYG23} as the baseline model, which is a binaural extension of the Waveformer model \cite{veluri2023real}. 

First, the input signal is processed by an encoder block:
\begin{equation}
    \mathbf{Y} = \text{Encoder}(\mathbf{x}),
\end{equation}
where $\text{Encoder}(\cdot)$ is the encoder layer, $\mathbf{Y} \in \mathbb{R}^{T' \times D}$ is the encoded representation of the input signal, $D$ is the feature dimension and $T'$ is the number of frames. 
The encoded mixture representation is then passed to an extractor, which predicts a mask $\mathbf{M} \in \mathbb{R}^{T' \times D}$ of the target sound to extract:
\begin{equation}
    \mathbf{M} = \text{Extractor}(\mathbf{Y}, \mathbf{e}^{\text{tgt}}),
\end{equation}
where $\text{Extractor}(\cdot)$ is an extractor neural network, and $\mathbf{e}^{\text{tgt}} \in \mathbb{R}^{D}$ is an embedding vector characterizing the target sound class. The embedding vector can be simply obtained from an embedding layer, which accepts the one-hot vector characterizing the desired sound class, $\mathbf{o}$, as input, i.e., $\mathbf{e}^{\text{tgt}} = \mathbf{W} \mathbf{o}$, where $\mathbf{W} \in \mathrm{R}^{D \times O}$ is an embedding matrix. 
Finally, the estimate of the target sound is obtained by applying the decoder to the masked features as 
\begin{equation}
    \mathbf{\hat{s}^{\text{tgt}}} = \text{Decoder}(\mathbf{M} \odot \mathbf{Y}),
\end{equation}
where $\text{Decoder}(\cdot)$ is the decoder layer, and $\odot$ is the Hadamard product.

Training the TSE system requires the triplets of binaural sound mixtures, target sound class, and binaural target sound signals. The optimal model parameters, $\hat{\theta}$ can be obtained by minimizing the training loss:
\begin{align}
    \hat{\theta} = \argmin_{\theta}    \mathcal{L}\left( \mathbf{s}^{\text{tgt}},\hat{\mathbf{s}}^{\text{tgt}}(\theta)\right).
\end{align}
Conventional TSE systems use only a signal-level training loss. Here, we investigate using a multi-task loss as:
\begin{equation}
    \mathcal{L} = \alpha \mathcal{L}^{\text{signal}} + \beta \mathcal{L}^{\text{spatial}}, \label{eq:comb}
\end{equation}
where $\alpha$ and $\beta$ are weights, and $\mathcal{L}^{\text{signal}}$, $\mathcal{L}^{\text{spatial}}$ are the signal-level and spatial losses, respectively, which we define below.

\subsection{Signal-Level Losses}
Conventional TSE systems are usually trained using a signal-level loss like the signal-to-noise ratio (SNR), the scale-invariant SNR (SI-SNR)\cite{le2019sdr} or a combination of both \cite{veluri2023real,VeluriICYG23}. We can use these losses for binaural outputs by computing a signal-level loss for each channel as proposed in \cite{VeluriICYG23}. For example, the SNR loss for binaural outputs is:
\begin{equation}
    \mathcal{L}^{\text{SNR}}(\mathbf{s}^{\text{tgt}},\hat{\mathbf{s}}^{\text{tgt}}) = - \left( \frac{1}{2}\text{SNR}(\mathbf{s}_{r}^{\text{tgt}}, \hat{\mathbf{s}}_{r}^{\text{tgt}}) + \frac{1}{2}\text{SNR}(\mathbf{s}_{l}^{\text{tgt}}, \hat{\mathbf{s}}_{l}^{\text{tgt}}) \right),
\end{equation}
where $\text{SNR}(\mathbf{s}, \hat{\mathbf{s}}) = 10\log_{10} \left( \frac{||\mathbf{s}||_2^2}{||\mathbf{s} - \hat{\mathbf{s}}||_2^2} \right)$ is the SNR between the reference signal, $\mathbf{s}$, and the estimated target signal, $\hat{\mathbf{s}}$. We can define a similar loss for the SI-SNR.

In this paper, following \cite{VeluriICYG23}, we use a weighted sum of SNR and SI-SNR losses as the signal-level loss:
\begin{equation}
    \mathcal{L}^{\text{signal}} = 0.9\mathcal{L}^{\text{SNR}} + 0.1\mathcal{L}^{\text{SI-SNR}}.
    \label{eq:sig_loss}
\end{equation}






\subsection{Spatial Losses}
To help the TSE model better preserve spatial cues, we investigate adding binaural metrics to the signal-level losses. For this purpose, we employ ILD and IPD losses, which were previously proposed for speech enhancement\cite{tokala2024binaural}, and also propose a new ITD-related loss. 

\subsubsection{Interaural Level Difference Loss}
ILD aims to measure the level difference between the channels of a signal. 
The ILD of a binaural signal is defined as:
\begin{equation}
    \text{ILD} = 10 \log_{10} \left( \frac{||\mathbf{s}_{ l}||_2^2} {||\mathbf{s}_{r}||_2^2} \right), \label{eq:ild}
\end{equation}
where $\mathbf{s}_{l} \in \mathrm{R}^T$ and $\mathbf{s}_{r} \in \mathrm{R}^T$ are left and right signals.

Then, the ILD loss can be expressed as the mean of the absolute difference of the target and predicted ILDs:
\begin{equation}
    \mathcal{L}^{\text{ILD}} =  \left| \text{ILD}^{\text{tgt}} - \widehat{\text{ILD}}^{\text{tgt}} \right|, \label{eq:ild_loss}
\end{equation}
where $\text{ILD}^{\text{tgt}}$ and $\widehat{\text{ILD}}^{\text{tgt}}$ are the ILD of the reference and extracted target sound signals computed with Eq. \eqref{eq:ild}. 
Note that prior work \cite{tokala2024binaural} defines this loss in the Short-Time Fourier Transform (STFT) domain, whereas we define it in the waveform domain.

\subsubsection{Interaural Phase Difference Loss} \label{sec:ipd}
The time difference of arrival (TDOA) between microphones translates into phase differences between the received sound signals. To preserve the spatial information of the sound source, one approach is to focus on maintaining the IPD of the reference left and right signals in the extracted signals.
The IPD between two signals can be calculated using the STFT of these signals.
However, the phase of a complex number has inherent periodicity (i.e., 0 and 2$\pi$ are equivalent). This property, known as the circularity problem, makes direct subtraction of phases unreliable. To address this, we compute the IPD between the two channels of a signal as:
\begin{equation}
    \text{IPD}_{u,v} = \text{atan} \left( \frac{\text{Im} \left( S_{u,v,l}  S_{u,v,r}^{*} \right)}{\text{Re} \left( S_{u,v,l}  S_{u,v,r}^{*} \right)} \right), \label{eq:ipd}
\end{equation}
where $S_{u,v,r}$, $S_{u,v,l}  \in \mathbb{C}$ represent the STFT of the right and left signals, $u$, $v$ are the indexes of the time and frequency bins, $^*$ is the conjugate operation, $\text{Im}(\cdot)$ and $\text{Re}(\cdot)$ represent the imaginary and real parts of a complex number, and $\text{atan}(\cdot)$ denotes the arctangent function. 

To leverage the IPD as a loss function, we compute the MSE between the IPD of the target and predicted signals:
\begin{equation}
    \mathcal{L}^{\text{IPD}} = \frac{1}{U V} \sum_{u=1}^U \sum_{v=1}^V  \left( \text{IPD}^{\text{tgt}}_{u,v} - \widehat{\text{IPD}}^{\text{tgt}}_{u,v} \right)^2,
    \label{eq:ipd_loss}
\end{equation}
where $\text{IPD}^{\text{tgt}}$ 
 and $\widehat{\text{IPD}}^{\text{tgt}}$ are the IPD of the reference and extracted target sound computed with Eq. \eqref{eq:ipd}, and $U$ and $V$ are the number of time and frequency bins, respectively.
 
Note that a similar IPD loss was used in \cite{tokala2024binaural}. They computed the IPD directly as the angle of the ratio $S_{u,v,l}/S_{u,v,r}$\footnote{\url{https://github.com/VikasTokala/BCCTN}}. This is mathematically equivalent to Eq. \eqref{eq:ipd} but handles the phase wrapping differently as Eq. \eqref{eq:ipd} defines the phase difference between $-\pi/2$ and $\pi/2$, i.e., the smallest possible phase difference. Besides, they used a binary mask to limit the IPD computation to the regions where the source is active. In our experiments, we did not use such a mask as it led to poorer performance, probably because of the difficulty of defining a mask suitable for the diversity of sounds covered by TSE.

\subsubsection{Proposed Interaural Time Difference Loss}
\label{subsec:prop}
Our aim is to develop a TSE system that can preserve the ITD of the target sound since it is an important spatial cue used by humans to localize sounds~\cite{blauert1997psychophysics,moore1991anatomy}. Therefore, we propose training the TSE model to output signals with an ITD close to that of the target sound.

ITD measures the difference in time of sound arrival between the left and right microphones. It can be obtained by finding the position of the highest peak in the cross-correlation between the left and right signals of the target sound. 
We can compute the ITD as:
\begin{equation}
    \text{ITD} = \argmax_{t \in [-\tau,\tau]} 
    \text{ } c_t, \label{eq:itd}
\end{equation}
where $c_t$ is the cross-correlation coefficient between the left and right signals at time step $t$.
$\tau$ is a scalar limiting the predicted delay to be in a given range, which is related to the distance between the microphones. 

Typically, the cross-correlation is computed using the generalized cross-correlation phase transform (GCC-PHAT) algorithm \cite{KrolikJPE84} as follows:
\begin{align}
&\mathbf{c} = \mathcal{F}^{-1}\left(\frac{\mathcal{F}(\mathbf{s}_{l}) \odot \mathcal{F}(\mathbf{s}_{r})^*}{\left| \mathcal{F}(\mathbf{s}_{l}) \odot \mathcal{F}(\mathbf{s}_{r})^* \right|}\right), \label{eq:gcc}
\end{align}
where $\mathbf{c} =[c_{t=-T}, \ldots, c_{t=0}, \ldots, c_{t=T} ] \in \mathbb{R}^{2T+1}$ is the vector of the cross-correlation coefficients, $\mathcal{F}$ and $\mathcal{F}^{-1}$ are the Fourier transform (FT) and inverse FT (IFT), respectively. We use the entire signal (here 6 seconds in both training and evaluation) as the windows length to compute $\mathcal{F}$.

A loss on the ITD should measure the difference between the ITD of the reference target sound signals, $\mathbf{s}^{\text{tgt}}$, and that of the extracted signals, $\hat{\mathbf{s}}^{\text{tgt}}$.
However, as seen in Eq. \eqref{eq:itd}, the ITD computation involves the argmax operation, which is not differentiable.
Therefore, we define the ITD loss as the mean squared error (MSE) between the cross-correlation of the reference and extracted signals as:
\begin{align}
    \mathcal{L}^{\text{ITD}} =  \frac{1}{2\tau + 1} \sum_{t = -au}^{\tau} \left( c^{\text{tgt}}_t - \hat{c}^{\text{tgt}}_t \right)^2
,
\end{align}
where $c_t^{\text{tgt}}$ and $\hat{c}_t^{\text{tgt}}$ are the cross-correlation between the reference and extracted signals computed with Eq. \eqref{eq:gcc}. Note that the proposed ITD loss is differentiable since all operations required to compute the cross-correlations, including FT and IFT, are differentiable.

\section{Experiments}
\label{sec:expe}

\begin{table*}[t]
\caption{signal-level and spatial metrics for mixture, baseline TSE using only spatial loss and proposed systems using signal-level and spatial losses. We report $\Delta$ITD-GCC values (using GCC-PHAT as in Eq. \eqref{eq:gcc}) and in parenthesis $\Delta$ITD values (using simple cross-correlation). }
\vspace{-3mm}
    \centering
    \begin{tabular}{
    @{}l|cc|ccc|c@{}
    }
    \toprule
        &  \multicolumn{2}{c|}{Signal-Level Metrics} & \multicolumn{3}{c|}{Spatial Metrics} & \\
          System & $\uparrow$ SI-SNR [dB] & $\uparrow$ SNR [dB] &  $\downarrow$ $\Delta$ILD [dB] & $\downarrow$ $\Delta$IPD [rad] &$\downarrow$ $\Delta$ITD-GCC ($\Delta$ITD) [$\mu$s] & $\downarrow$ FR [\%]\\
         \midrule
         (1) \quad Mixture & -0.74 & -0.73 & 2.68 & 0.84 & 235.7 (263.0) &- \\
         (2) \quad Baseline TSE w/ $\mathcal{L}^{\text{signal}}$ & 6.50 & 7.85 & 0.84 & 0.88 & 163.5 (86.3) & 0.17 \\
         \cmidrule{1-7}
         (3) \quad (2) + $\mathcal{L}^{\text{ILD}}$ &  6.72 & 8.10 & $\textbf{0.74}$ & 0.83 & 168.5 ($\mathbf{74.8}$) & \textbf{0.16} \\ 
         (4) \quad (2) + $\mathcal{L}^{\text{IPD}}$ &  \textbf{6.76} & 8.03 & 0.79 & $\textbf{0.49}$ & 242.9 (80.1) & \textbf{0.16} \\ 
         (5) \quad (2) + $\mathcal{L}^{\text{ITD}}$ & 6.74 & $\textbf{8.11}$ & 0.78 & 0.84 & $\textbf{137.3}$ (79.0) & \textbf{0.16} \\ 
        \bottomrule
    \end{tabular}
    \label{tab:results}
    \vspace{-4mm}
\end{table*}

\subsection{Experimental Settings}
\subsubsection{Datasets}
In our experiments, we used an openly available dataset of binaural sound mixtures \cite{VeluriICYG23}. The data consists of simulated reverberant mixtures of three to four sound events added to urban background noise. 
This dataset leverages 20 sound classes from FSD50K \cite{FonsecaFPFS22} (general-purpose), ESC-50 \cite{Piczak15} (environmental sounds), MUSDB18 \cite{musdb18} and noise files for the DISCO dataset~\cite{furnon2021dnn}. 
The background sounds were taken from TAU Urban Acoustic Scenes 2019 \cite{MesarosHV18}. Binaural mixtures were generated by convolving the sound source signals with room impulse responses (RIRs) and head-related transfer functions (HRTFs) as it is done in Semantic Hearing \cite{VeluriICYG23}. We used HRTFS from 43 subjects from the CIPIC corpus\cite{algazi2001cipic}, and real and simulated RIRs from three corpora\cite{SBSBRIR,IoSR_Surrey_2016,IoSR_Surrey_2023}. The sampling frequency of all signals was 44.1 kHz.

We mixed the data on the fly with Scaper \cite{salamon2017scaper}. From each mixture, we only extracted one of the foreground sources (the same one across the experiments) \cite{VeluriICYG23}. 
The training was done with 6-second mixtures. The number of training, validation, and testing mixtures were 100K, 1K, and 10K, respectively.

\subsubsection{System Configuration}
We employed the same configuration for the TSE system as used in \cite{VeluriICYG23}.\footnote{We used the model variant $D = 256$ proposed by Semantic Hearing  \url{https://github.com/vb000/SemanticHearing}}  The system performs online processing with a latency of 20 msec. It has  1.74M trainable parameters in total.

The model consists of a 1-D convolution and 1-D transposed convolution for the encoder and decoder, respectively. The encoder  1-D convolution layer has a stride of $L=32$ samples and a kernel size of $K=3L$. 
After the mask is obtained with the extractor, the 1-D transposed convolution layer returns the waveform of the target sound.
The extractor consists of an encoder-decoder architecture conditioned on the target class embedding by multiplying the output of the dilated convolution with the target sound embedding, $\mathbf{e}^{\text{tgt}}$. The encoder is a stack of 1-D dilated convolution (DCC) layers followed by a Transformer-like layer. The DCC layers have a kernel size of $K$=3 and dilation factors set to $\{ 2^0, 2^1, ..., 2^9 \}$. The Transformer-like layer consists of two multi-head attention (MHA) layers with 8 heads. 

We trained all models for 80 epochs using a batch size of 32 and a learning rate of 5e-4. 
For the IPD loss, we computed the complex STFT with a Hann window of 1024, a hop length of 256, and an FFT of 1024 samples. For all models, we used the signal-level loss defined in Eq.~\eqref{eq:sig_loss}, and combined it with the spatial losses according to Eq.~\eqref{eq:comb} with fixed weight $\alpha=1$. 
We selected the hyperparameters, including the loss weights $\beta$ and the number of epochs, that achieved the best spatial metric among those that did not degrade signal-level metrics on the validation set. Accordingly, the values of $\beta$ were chosen to 0.1, 1, and 1 for the ILD, IPD, and ITD losses, respectively. For ITD loss, we set $\tau = 1$ms. 

\subsubsection{Evaluation Metrics}
Following Semantic Hearing \cite{VeluriICYG23}, we measured the performance on both signal-level (SI-SNR and SNR), and spatial metrics by computing the difference between the reference and extracted signals, i.e., $\Delta$ILD, $\Delta$IPD, and $\Delta$ITD.  $\Delta$ILD and $\Delta$IPD were computed with Eq.~\eqref{eq:ild_loss} and Eq.~\eqref{eq:ipd_loss}, respectively. $\Delta$ITD was obtained as the absolute difference between the ITD of the reference and extracted signals, where ITD is obtained with Eq.~\eqref{eq:itd}. To compute ITD, we used the GCC-PHAT as it is known to be more robust to reverberation. However, we also report results with the simple cross-correlation for comparison with prior work \cite{VeluriICYG23}. We name these metrics $\Delta$ITD-GCC and $\Delta$ITD, respectively. 
Note that prior works designed spatial metrics considering human perception \cite{dietz2011auditory,blauert1997psychophysics}, by, e.g., computing ITD only on the frequencies below 1.5 kHZ and ILD on the frequencies above 3 kHz. With such evaluation metrics, we observed a small improvement over the baseline of about 5 $\mu$s with our proposed ITD loss. However, we expect a larger improvement if we would include a low-pass filter in the loss computation, which will be part of future investigations.



Finally, we also report the failure rate (FR) to provide a rough estimate of how often TSE failed to correctly identify the target sound in the mixture. We defined FR as the percentage of test samples for which SI-SNR improvement is below 1dB \cite{delcroix22_interspeech}.  

\subsection{Experimental Results}
Table~\ref{tab:results} compares the performance of the proposed losses with that of the unprocessed mixture and a baseline TSE that only uses the signal-level loss of Eq. \eqref{eq:sig_loss} \cite{VeluriICYG23}. Note that the ILD, ITD and, IPD values computed on the mixture are only provided as a reference. We observe that the baseline TSE greatly reduces ILD and ITD errors compared to the mixture but not IPD. Actually, most experiments failed to improve IPD errors, which may suggest the difficulty of using IPD metrics for TSE.


Using the spatial losses in addition to the signal-level loss (systems (3)-(5)) outperforms the baseline TSE system (system (2)) for most metrics. These results show that the multi-task loss does not degrade extraction performance in terms of signal-level metrics and FR (it even slightly improves it).

We confirm that adding a spatial loss improves performance on the related spatial metric.
Using the ILD loss (system (3)) performs best in terms of $\Delta$ILD but does not consistently improve $\Delta$ITD. 
Using the IPD loss (system (4)) greatly reduces $\Delta$IPD, and still slightly improves $\Delta$ILD. However, it does not translate into improving ITD errors, suggesting the limit of the IPD loss.
In contrast, using our newly proposed ITD loss (system (5)) improves all spatial metrics compared to the baseline while achieving comparable SI-SNR, SNR, and FR values. The improvement in terms of $\Delta$ITD is particularly large, i.e., 26.2$\mu$s (more than 1 sample) or a relative improvement of 16 \%. These results confirm that our proposed ITD loss contributes to improving spatial cue preservation of TSE.


\section{Conclusions}\label{sec:conclusion}
In this paper, we introduced a new loss function based on ITD, which preserves binaural properties and improves ITD difference between left and right channels. Our proposed ITD loss is superior to a conventional IPD loss as it improves binaural cue recovery of TSE in terms of ILD, IPD, and ITD, while maintaining performance in terms of signal-level metrics. 

The proposed ITD loss is general and could be used for other speech and audio processing tasks, such as binaural speech enhancement or speech separation. Future works will include such investigations.

\bibliographystyle{IEEEbib}
\bibliography{refs}

\end{document}